\documentclass[12pt]{article}

\usepackage{graphicx}
\usepackage{scicite}

\usepackage{bm}

\usepackage{times}

\newenvironment{sciabstract}{%
\begin{quote} \bf}
{\end{quote}}

\newcounter{lastnote}

\title{A bright point source of ultrashort hard x-rays from laser bioplasmas}

\author
{M. Krishnamurthy$^\ast$, Sudipta Mondal, Amit D. Lad, Saima  Ahmad, V. Narayanan,\\
R. Rajeev, M. Kundu$^\top$, G. Ravindra Kumar and  Krishanu Ray \\
\\
\normalsize{Tata Institute of Fundamental Research, Homi Bhabha Road, Mumbai, 400 005, India.}\\
\normalsize{$^\top$Institute for Plasma Research, Bhat, Gandhinagar, 382 428, India }\\
\\
\normalsize{$^\ast$To whom correspondence should be addressed; E-mail:  mkrism@tifr.res.in}
}

\date{}

\begin{document} 


\baselineskip24pt


\maketitle


\begin{sciabstract}
  Intense, femtosecond laser created plasmas\cite{review} are unique tabletop point-sources of pulsed energetic  electrons\cite{malka}, ions\cite{proton} and x-rays with energies\cite{droplet} extending well into the MeV region. These emissions, originating from preferentially heated plasma electrons\cite{gibbon} are of great importance to diverse scientific, technological and medical applications such as radiography\cite{radiography}, cancer therapy\cite{proton} and  lithography\cite{lithography}. Nanoscale structuring of the target surface by particle deposition\cite{rajeev} or sub-lambda gratings\cite{kahaly} enhances these emissions {\it via} local field amplification caused by scattering and/or surface plasmon coupling. Such artificial target preparation, however, suffer from limitations in the range of structures  achievable, apart from the restrictions imposed by the manufacturing process\cite{nanomethods}.  Besides, laser intensities required for hot electron generation using such targets\cite{xray-efficiency}  are currently large enough to limit the source repetition rate to $<$100 Hz. Here we demonstrate a novel  structured target that liberates these sources from the above constraints. A target consisting of a few micron-thick layer of an ubiquitous microbe, {\it Escherichia coli} (E. coli), catapults the brightness of hard x-ray  bremsstrahlung emission (up to 300 keV) by more than 100 fold at  an incident laser intensity of 10$^{16}$ W cm$^{-2}$.  This increased yield is attributed to the local enhancement of electric fields around individual {\it E. coli} cells and  is reproduced by detailed particle-in-cell (PIC)\cite{pic} simulations. This combination of laser plasmas and biological targets can  lead to turnkey, multi-kilohertz and environmentally safe sources of hard x-rays. It would also trigger exploration of an unlimited diversity of biological materials in nature as targets for laser plasma generation. 

\end{sciabstract}




Intense ultrashort laser pulses deliver large amounts of energy to matter in time scales too short for the atoms in the target to move significantly. At 10$^{16}$ W cm$^{-2}$, 100 fs pulses are known to transfer up to 240 mW/atom  in nanoclusters of a few thousand atoms\cite{rhodes}. The energy absorption is mediated by the ionization of the atoms and heating of the electrons. Such large energy absorption results in fully stripping all the electrons of atoms like Argon\cite{rhodes}. With a solid slab target, the plasma electron temperature reaches a few tens to hundreds of keV\cite{batani} and results in x-ray emission both in a continuous spectrum due to bremsstrahlung and characteristic lines spectra due to the inner shell transitions of the atomic constituents. Since the hot electrons are generated by femtosecond bursts of light, the corresponding x-ray emission lasts for a similar duration\cite{picoxray}. 

Hot electrons are generated by a number of processes: a) resonance absorption (RA), b) vacuum heating, c) J $\times$ B heating\cite{gibbon} {\it etc.,} of relevance to our discussion here is RA, which occurs when the strong laser field drives plasma electron waves, which  damp out to produce hot electrons\cite{sandhu}.  RA is excited only by p-polarized light that is incident at an angle, (typically 30-45 degrees) to the normal of the target. Experimental and simulation evidence indicates that RA occurs  efficiently above 10$^{15}$ W cm$^{-2}$  and about 30\% of the laser light can be coupled to the hot electron generation\cite{gibbon}. The hot electron energies are characterized by well established scaling laws, such as the Forslund scaling law\cite{Forslund}, which yields, T $_{hot}$ (keV) = 14 (T$_c$ I $\lambda^2$)$^{1/3}$, where  T$_c$  is the  bulk plasma temperature in keV,  I is the laser intensity in units of 10$^{16}$ W cm$^{-2}$ and  $\lambda$ is the laser wavelength in $\mu$m. 

A method that enhances the absorption of the laser energy and therefore, the ability to increase hot electron/x-ray emission, is of paramount importance. This has been addressed in two ways. The first  relies on optimizing laser parameters such as intensity, pulse width, polarization and temporal shape. A second aims to engineer the target  for efficient absorption and subsequent channeling of the absorbed radiation to increase a specific process, x-ray emission,  for example. This latter approach has just begun to be realized \cite{rajeev,kahaly}. For instance, a 15 $\mu$m liquid droplet is shown to produce two orders of magnitude amplification in the local fields and 60 fold enhacement in the hard x-ray yield\cite{droplet}. Similarily, a velvet target\cite{majoribanks} with complex nanostructuring yields 50-fold enhacement in  the soft x-ray emission. In both cases the complexity of target preparation and experimental procedures involved in maintaining the target in vacuum are major limitations. Forslund scaling law\cite{Forslund} clearly implies that enhanced intensities will lead to hotter electrons. Scattering of light in nano- or micro-structures of the target can amplify  local electric fields and the enhanced local light intensity  drives the plasma more effcicently. It is this aspect that we exploit in the present experiments by using micron sized bacterial cells, namely {\it E. coli}. 

{\it E. coli} cells, which constitutes the targets in this work, are ellipsoidal in shape ($\approx$ 1.8 $\mu$m on major axis and $\approx$ 0.7$\mu$m on minor axis)\cite{bacsize}. Each cell also expresses numerous flagella, 25 nm in diameter and 10-20 $\mu$m in length from their cell exterior\cite{bacsize}.  In the present experiments, we used a laboratory strain of {\it E. coli} DH5alpha and the cells were chemically fixed before target preparation. The fixation process sheds the flagella and therefore only the ellipsoid structure of the cells was exploited. It has no known subcellular structures. Hence, we assume that the majority of the 0.7 femtolitre volume is filled with optically uniform cytoplasm.  For the purpose of this work, these cells can also be viewed as microparticles with well defined sizes that are filled with low z- atoms. 

A conceptual picture of the novel hard x-ray source is given in Figure 1. A femtosecond laser source (40 fs, 800 nm) is focused at a 45$^\circ$ incident angle to a 17$\mu$m spot (intensity of $\approx$10$^{16}$ Wcm$^{-2}$) on  a solid glass plate coated with a few microns thick layer of {\it E. coli} cells.  The cells are `instantaneously' (in a few femtosecond) converted into a dense, hot plasma  that  further absorbs the laser light  and finally radiates its energy in the form of x-ray pulses ranging up to 300 keV.   The effect of the bacterial coating on x-ray emission is deciphered by comparison with the emission from an uncoated portion of the glass surface under identical conditions.  Figure 2a shows the hard x-ray spectrum measured using a calibrated NaI(Tl) scintillation detector.  As seen in this figure, the hard x-ray emission  is very high with the bacterial coating (blue solid circles), while it is hardly visible for an uncoated glass (shown in pink triangles) under similar conditions. The total yield of hard x-rays integrated over 50-300 keV, is about 120 times larger in the former case. We carried out similar experiments under identical conditions with homogenised bacteria  (shown in squares) to prove that the effect is primarily due to the light scattering from the shape of bacteria (the target coating is shown in Figure 2b) and the associated local fields.  Most of the bacterial cells were disrupted by the homogenisation  process (Figure 2c), and would thus eliminate the local field enhancement and thereby reduce the x-ray yield.  This is indeed observed: the homogenised bacterial layer of similar thickness produced x-rays with much diminished efficiency, the total emission being mariginally larger than that observed with glass slab (squares, Figure 2a). Inset in Figure 2a shows a Maxwell-Boltzmann fit (shown as solid line)  to the x-ray emission from the bacterial coated target. The electron temperature from the fit is about 57 $\pm$ 2 keV  for the bacterial coated target,  about 2.5 times larger than that from an uncoated glass target under identical conditions. 

Clearly, these data indicate that the incident light interacts more efficiently with the microstructures, the plasma is hotter and the x-ray emission is brighter in the coated targets. This enhanced x-ray emission can easily be explained by invoking local electric field and intensity enhancement\cite{Boyd, rajeev, rajeev-ol}. The enhanced local fields would create plasma hot spots,  which would further increase the absorption of laser light and the electron generation\cite{droplet}.  Two dimensional particle-in-cell (PIC) simulations, presented below, provide compelling  support to the proposed role of the microstructures in hot electron generation.    

 Light scattering from ellipsoid shaped plasma structures on the slab  modifies the local intensity, computation of effective intensity using 2D-PIC is  shown in Figure 3a and we observe hot spots that are about 5 times more intense than the incident light.   These hot spots would further influence the laser absorption and plasma generation. The hot electron spectrum from PIC simulations give the overall effect of these hot spots and these results are plotted in Figure 3b. The black solid squares show the electron spectrum for only an uncoated slab, which is devoid of these enhanced local fields and would be akin to an uncoated glass substrate. The hot electron yield and temperature increases with the addition of ellipsoid structures on the solid slab. We have done calculations with different number of ellipsoids on the solid slab as shown in the Figure 3b. The real experiment has more than a monolayer of bacteria spread throughout the focal spot. The simulations with increasingly larger number of ellipses were carried out to approach the conditions closer to the experiment and to find out how the hot electron yield changes in such situations. A larger number of ellipsoid particles increases the hotter electron component as can be seen from plots for one, three and eleven ellipses in Figure 3b. It shows that the number of hot spots increases with the increase in ellipses and the hot electron population would increase likewise.  

Hot electron spectra generated from the 2D-PIC simulations fits well with a two temperature distribution shown as solid lines in Figure 3b. For the solid slab (black squares) the electron temperatures are about 5$\pm$1 keV with an insignificant higher temperature component.  With eleven ellipses, nearly the entire focal area is filled with the bacterial cells and therefore, is more comparable to the experimental conditions. The hotter second temperature for this (red circles) is about 70 $\pm$2 keV, which is reasonably close to the experimental measurements (57 keV observed with bacterial coating). We also observed that the hotter electron temperature in simulations is also about 2.5 times that of the uncoated target, close to the experimental measurements. Further, while the temperature remains similar with the larger number of ellipsoids,  the hot electron yield  increases very effectively. The inset in Figure 3b shows the hot electron yield above 50 keV for different number of ellipses on the slab. For eleven ellipses the hotter electron yield is about 50 times that of the solid slab. In the experiments, however, the number of ellipses was more and thus the enhancement obtained was about 120 fold. Since the hot electron component increases with the number of ellipsoids, addition of monolayers of ellipsoids is perhaps required in the simulations to quantitatively reproduce the experimentally measured enhancement. Even with the simplicity presumed in the computations and the modeling of the bacterial cells as ellipsoids, the simulations very effectively reproduce the enhancement in the hotter electron yield as measured by the x-ray spectrum. 

It is important to note that a new paradigm in x-ray generation from intense laser matter interaction has been initiated through this work. A number of new questions of both technological and fundamental interest arise from these experiments. For example, the role of the atomic constituents of the cell. If the bacterial cells are doped with high z- materials it is very likely that the x-ray emission can be further boosted. Our preliminary results  are already encouraging. It is also possible to align the bacterial cells to manipulate the light scattering, the local fields as well as the x-ray generation. The smoothness of the cell wall or the presence of nanostructure on the cell surface could also affect the plasma evolution. We have shown that  the local fields play a very crucial role in plasma generation using a simple shape of {\it E. coli}. This indicates that biological cells with well ordained microstructures can be exploited as effective targets for hotter plasma generation and a resultant bright point-source of radiation with energy all the way upto the hard x-ray regime (a few hundred keV). Natural cells provide a myriad of micro structures that can be well exploited for this purpose. Apart from the coccus (spherical), spirillum (spiral) or filamentous bacteria even more complex shapes due to stalks and appendages in species like caulobacter, Myxococcus or Streptomyces can be exploited\cite{bacshapes}. The use of the right shape can deliver optimal enhancement of the local fields and x-ray emission.

\noindent {\bf \Large  Methods}

\noindent {\bf \large  Sample preparation and analysis}

A regular strain of Escherichia coli bacteria was grown overnight in a suspension culture in minimal media. We used both live or chemically fixed and UV attenuated cells. Though the effect is similar, most of the experiments were carried out with fixed and attenutaed cells to avoid generation of any virulent strains due to radiation exposure. Fixation is carried out by the use of a mixture of 4$\%$ formaldehyde and 2.5 $\%$ gluteraldehyde solutions. To coat bacteria on a optically polished BK-7 glass substrate, we  first painted it with 1 mg/ml poly-L-lysine solution and air dried for a few minutes. This creates charged hydrophilic surface on the substrate and helps to form a uniform coating of the bacterial cells. The cell-suspension was then applied on the ploy-L-lusine coated glass, alloewed to dry in a laminar flow hood and  irradiated in a suitable chamber with appropriate UV dose (250 mJ of 280-300 nm UV) to fully attenuate allthe bacterial cells. The coated target slabs were then left to dry in a desiccator. Thus fixed, the {\it E. coli} cells remained structurally unaffected under  10$^{-5}$ Torr vacuum for over an hour. A profilometer scan of the the bacterial coatings showed that the it was fairly uniform with height variations between 1-2 $\mu$m, which is equivalant to 2-3 layers of { \it E. coil} cells. The bacterial homogenate was prepared by sonication of the { \it E. coli} suspension for a several minutes and then  the emulsion was coated on the glass surface in an identical manner as described above. 

\noindent {\bf \large Experimental setup} 
 
 The experiment was conducted with a  40 fs / 10 Hz, 20 TW Ti:Sapphire laser system at Tata Institute of Fundamental Research, Mumbai. In the present experiments, the laser energy is  restricted to less than 10 mJ on target and the laser pulses are focused with an f/3 off-axis parabolic mirror to a 17 $\mu$m focal spot diameter, giving a maximum focal intensity of 5 $\times$ 10$^{16}$ W/cm$^2$. The target is moved with a stepper motor controlled system such that every laser pulse is incident on a fresh unused portion of the target. A photo-diode is used to monitor the pulse energy  fluctuations. A 5 cm $\times$ 5 cm target can produce 20000-50000 x-ray pulses.   The x-ray measurements are carried out with a 2 inch thick NaI (Tl) scintillation detector coupled with conventional electronics that include a pre-amplifier, amplifier, pulse-height discriminator and multi-channel analyser for computerised data acquisition.  The detector system is gated with a 20 $\mu$s gate pulse generated in synchronous to the laser trigger. The detector is placed at about 70 cm from the target across a 5 mm glass window. Pile-up free detection is ensured by shielding the detector in a 10 mm thick Lead housing with an appropriate size aperture such that the count rate is less that 1 count in 10 pulses. Under otherwise identical condition the x-ray energy and yield are measured for over 10,000 laser shots by translating the target on both the bacterial coated portion and uncoated portion to make an objective comparison. 
 
\noindent {\bf \large Computational Methodology} 
 
 The two dimensional particle-in-cell simulation are carried out on a 1000$\times$1000 grid with uniform grid size $\Delta=\lambda/40$ to simulate the observations.  A solid slab of 20$\lambda\times$5$\lambda$ with ellipsoid particles of similar size as {\it E. coli} (0.7$\mu$m$\times$ 1.8$\mu$) is illuminated by normally incident light of $\lambda$ = 800 nm and  intensity 10$^{16}$ W cm$^{-2}$ to simulate conditions close the experiment. Assuming slowly varying envelope approximation\cite{mporasPRE2002, yanAppB2005}, a continuous Gaussian beam, with a transverse field component $E_x(t,x,y=y_b)=E_0 \left(w_0/w(y)\right)\exp(-r^2/w(y)^2)\mathrm{Re}[\exp(\mathrm{i} \omega (t - y/c) + \mathrm{i}\arctan(y/y_R) - \mathrm{i} \omega r^2/2 c R(y))]$ (focal width $w_0 = 20\lambda$, Rayleigh range $y_R = \omega w_0^2/2c$, radius of curvature $R(y) = y + y_R^2/y$) is numerically excited at one end $y=y_b$ of the computational box. It is propagated so that the entire structure can be uniformly illuminated. The focal spot accomodates an increasing number of ellipsoid structures in front of the solid slab, keeping the interaction time $\approx 35$~fs and then switched off. A uniform initial electron density $2 n_c$ is assumed for slab and ellipses, where $n_c$ is the critical electron density ($\sim 1.72\times$10$^{21}$ cm$^{-3}$ at 800 nm). 
  
 \noindent {\bf \large Acknowledgement} 
We thank J. Seth for the help in initial set of experiments, L. Borde and S. Shirolikar of the TEM facility for electron microscopic images and A. Venugopal for profilometer measurements. We also thank C. Danani and TBM group of IPR for executing simulations in their workstations. MK  thanks DST, India for  Swarnajayanti Fellowship grant and GRK acknowledges a DAE-SRC-ORI grant.
 \clearpage


\begin{figure}
\includegraphics[width=1.0\textwidth]{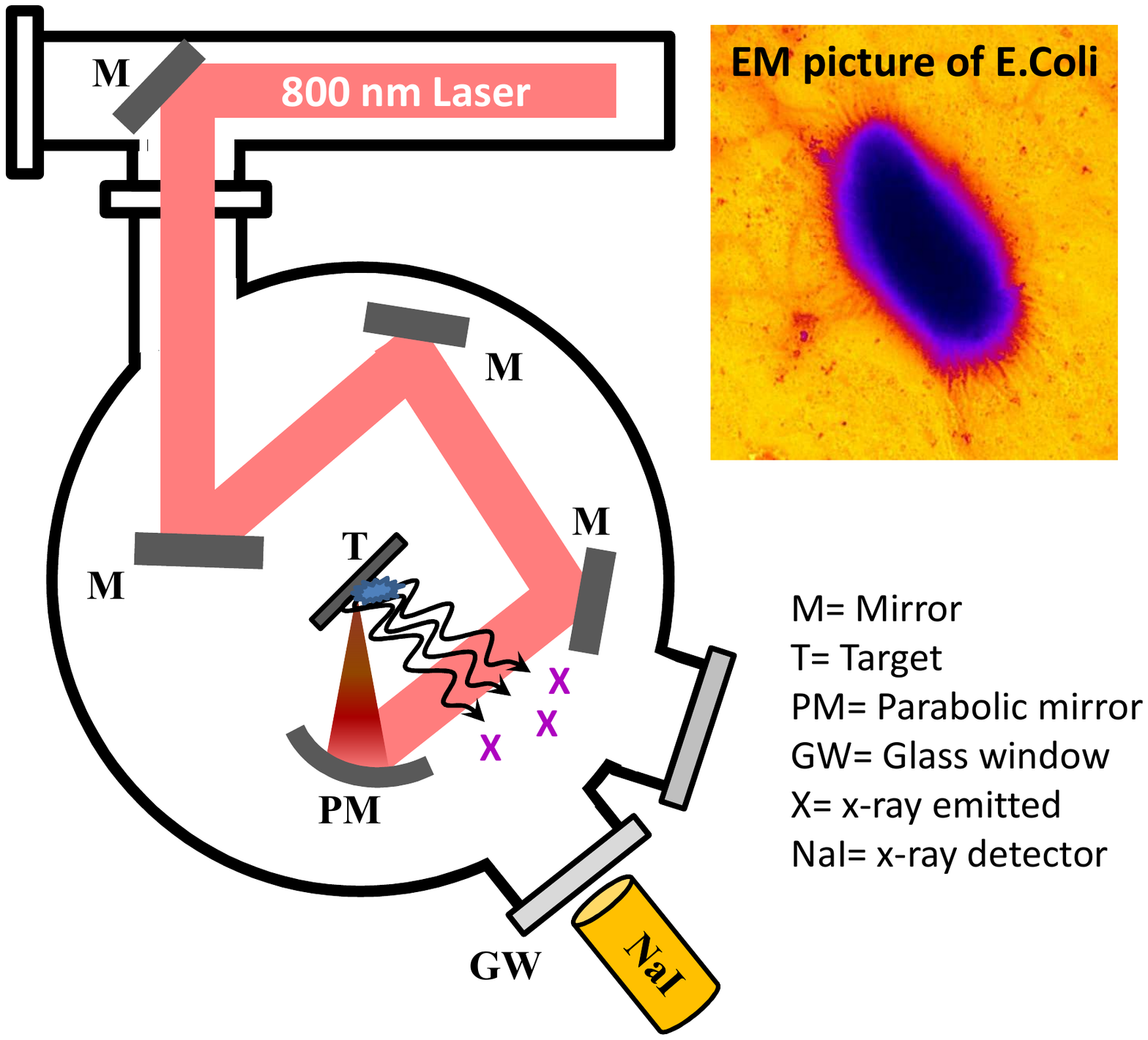}
\vskip -3cm
\caption{Schematic of the experimental apparatus. A 40 fs pulse of 800 nm light is focused on the target to intensities of about 10$^{16}$ W cm$^{-2}$ using a off-axis parabolic mirror. The target is glass substrate coated with a few micron layers of {\it E. coli} bacteria. Inset shows an electron microscope (EM) image of {\it E. coli}.
}
\end{figure}

\begin{figure}
\vskip -3cm
\includegraphics[width=1.0\textwidth]{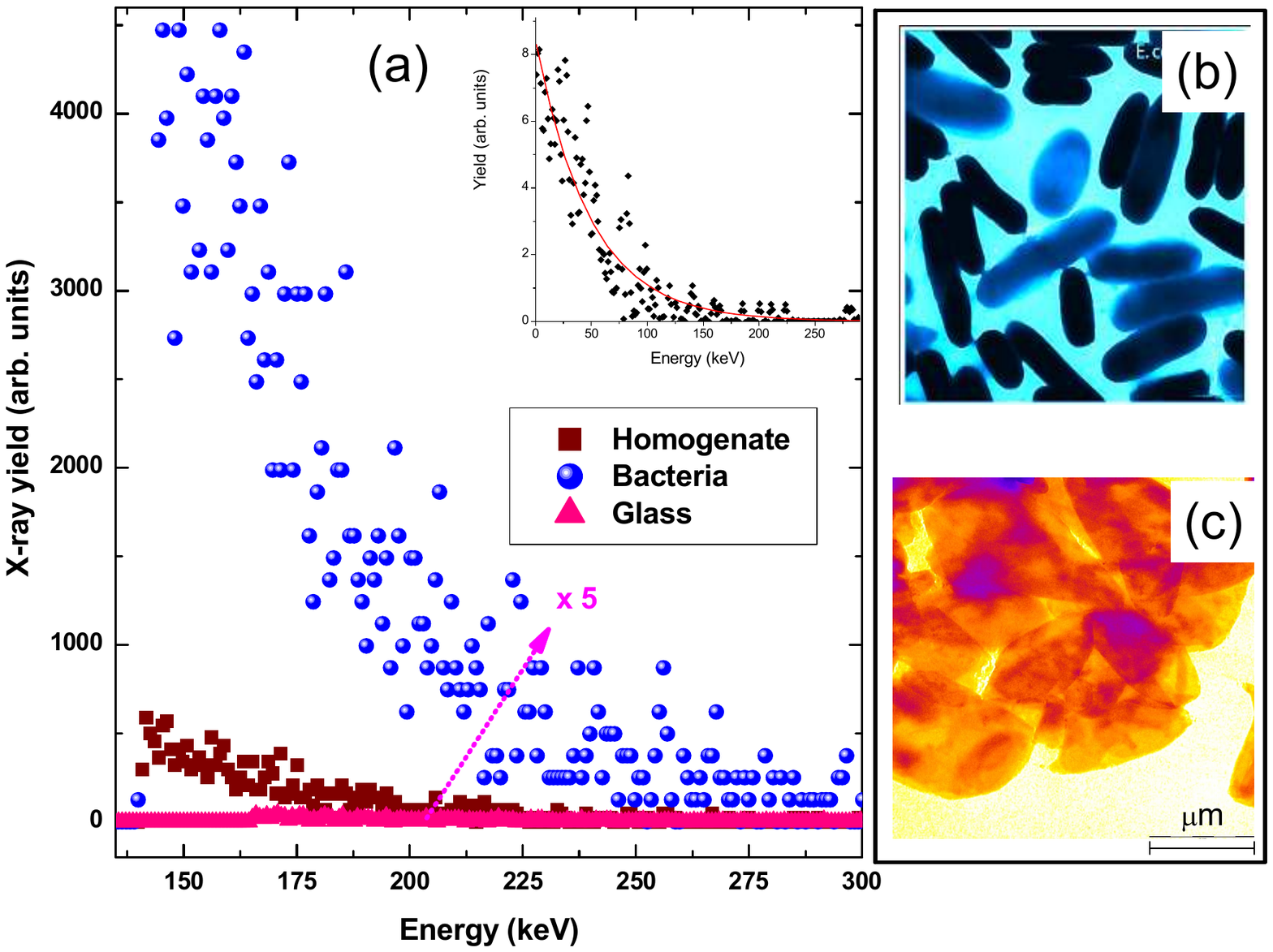}
\vskip -7cm
\caption{Bremsstrahlung x-ray spectrum up to about 300 keV, measured  for different types of targets  at a laser intensity of 5 $\times$ 10$^{16}$ W cm$^{-2}$.   The topmost curve (blue circles) represents the yield from  the {\it E. coli} coated target and  it is evident that this is significantly larger than the emissions from the uncoated (pink triangles) and  homogenate coated targets (magenta squares).  The integrated yield from the {\it E. coli} coated target is about 120 times larger than that from the uncoated glass. The homogenate is show only 23 times larger yield. The inset shows the x-ray spectrum from {\it E. coli} over the entire energy range and the solid line shows an exponential fit with an electron temperature of 57$\pm$2 keV. A false color EM images of the intact {\it E. coli} cells is shown in  (b) and the homogenate is shown in (c). }.

\end{figure}

\begin{figure}
\vskip -5cm
\includegraphics[width=1.0\textwidth]{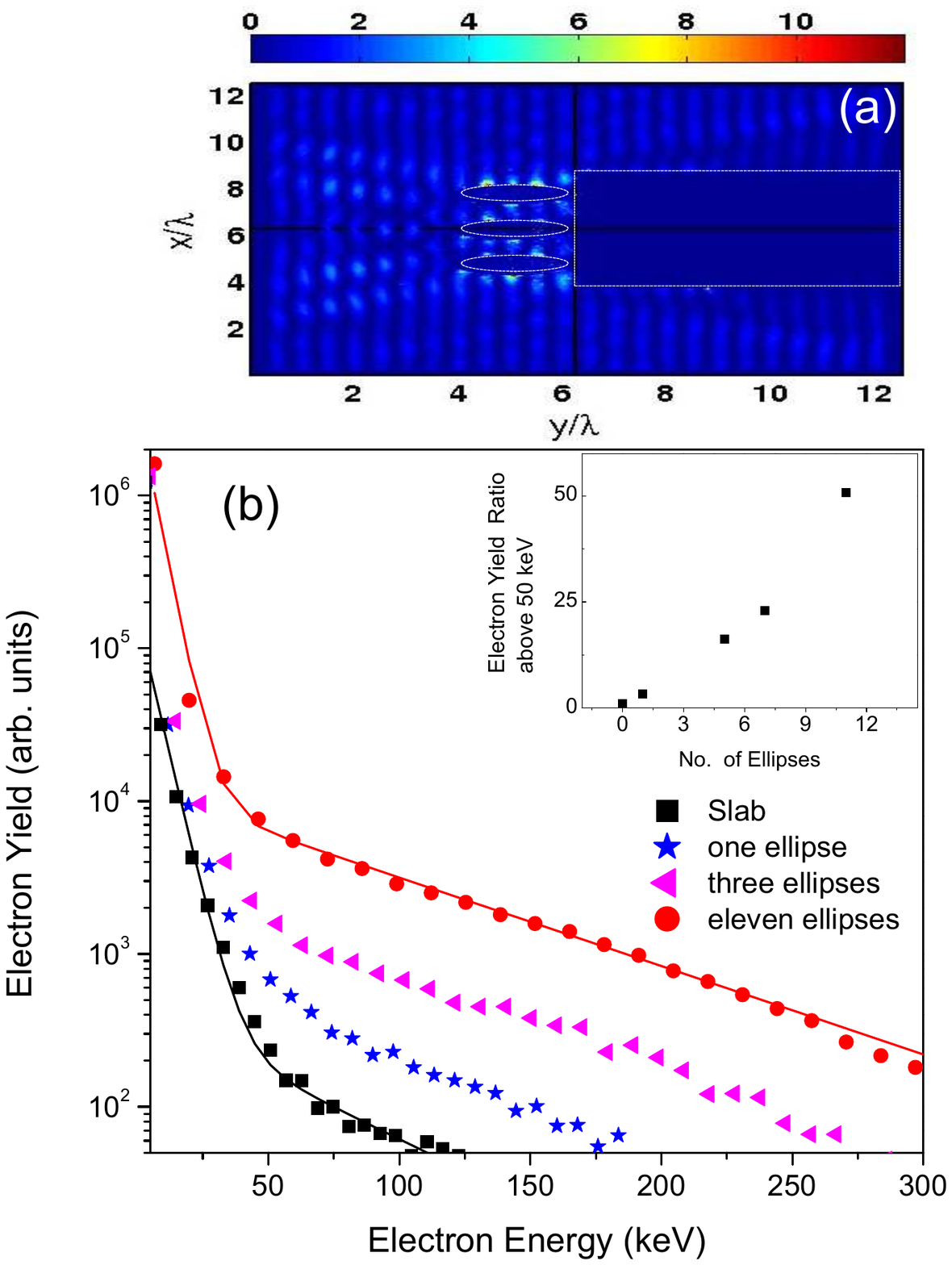}
\vskip -4cm
\caption{ 
a) Light intensity pattern simulated by 2D-PIC when a solid slab coated with three ellipsoid particles of 0.7$\mu$m $\times$ 1.8$\mu$m is irradiated with 35 fs pulses of 10$^{16}$ W cm$^{-2}$. The while dotted line shows the initial boundary of the solid slab and the ellipsoid particles. b) Electron spectrum derived from the 2D-PIC simulations with varied number of ellipsoid particle ion the solid as indicated in the legend. Slab refers (black square) refers to electron spectrum with solid slab and eleven refers (red circles) to calculation with eleven ellipsoid particles on the solid slab. The solid line show a two temperature exponential fit to the simulated data. See the text for details. Inset in (b) shows the hot electron yield above 50 keV for calculations with different number of ellipses. The solid slab yield is normalised to one to obtain a relative enhancement.
}
\end{figure}
\end{document}